\documentclass[pdflatex,sn-mathphys]{sn-jnl}
\usepackage{amsmath,amsthm,mathtools}
\usepackage{amssymb}
\usepackage[utf8]{inputenc}
\usepackage{upgreek}
\usepackage{xcolor}
\usepackage{soul}

\usepackage{lineno}

\begin{document}

\title{Quantum Vortex Formation in the \textit{``Rotating Bucket''} Experiment with Polariton Condensates}

\author[1]{\fnm{Ivan} \sur{Gnusov}}

\author[2]{\fnm{Stella} \sur{Harrison}}

\author[1]{\fnm{Sergey} \sur{Alyatkin}}

\author[1]{\fnm{Kirill} \sur{Sitnik}}

\author[1]{\fnm{Julian} \sur{T\"{o}pfer}}

\author[2,3]{\fnm{Helgi} \sur{Sigurdsson}}

\author*[1,2]{\fnm{Pavlos} \sur{Lagoudakis}}\email{Pavlos.Lagoudakis@soton.ac.uk}

\affil[1]{\orgname{Hybrid Photonics Laboratory, Skolkovo Institute of Science and Technology}, \orgaddress{\street{Territory of Innovation Center Skolkovo, Bolshoy Boulevard 30, building 1}, \postcode{121205} \city{Moscow}, \country{Russia}}}

\affil[2]{\orgdiv{School of Physics and Astronomy}, \orgname{ University of Southampton}, \orgaddress{\city{Southampton}, \postcode{SO17 1BJ}, \country{UK}}}

\affil[3]{Science Institute, University of Iceland, Dunhagi 3, IS-107, Reykjavik, Iceland}

\abstract{The appearance of quantised vortices in the classical ``rotating bucket'' experiments of liquid helium and ultracold dilute gases provides the means for fundamental and comparative studies of different superfluids. Here, we realize the ``rotating bucket'' experiment for optically trapped quantum fluid of light based on exciton-polariton Bose-Einstein condensate in semiconductor microcavity. We utilise the beating note of two frequency-stabilized single-mode lasers to generate an asymmetric time-periodic rotating, non-resonant excitation profile that both injects and stirs the condensate through its interaction with a background exciton reservoir. The pump-induced external rotation of the condensate results in the appearance of a co-rotating quantised vortex. We investigate the rotation-frequency dependence  and reveal the range of stirring frequencies (from 1 to 4 GHz) which favors quantised vortex formation. We describe the phenomenology using the generalised Gross-Pitaevskii equation. Our results enable the study of polariton superfluidity on a par with other superfluids, as well as deterministic, all-optical control over structured nonlinear light.

}

\maketitle

 \section*{Introduction}\label{sec1}

Orbital angular momentum (OAM) in paraxial light, or optical vorticity, is an essential degree of freedom, alongside polarization, for optical information encoding and processing~\cite{30yearsoptical}. This has sparked a strong interest in developing microlasing devices emitting a rotating phase front of the electromagnetic field of controlled OAM~\cite{Jincheng_Science2021}. Being almost strictly noninteracting systems, optical vortices differ dramatically from conventional vortices in interacting fluids. The latter are ubiquitous in nature ranging from the enormous vortex storms in the gas belts of Jupiter~\cite{jupiter} to tiny micrometer-size quantum vortices in interacting macroscopic quantum systems such as superconductors~\cite{superconductor}, superfluids~\cite{donnelly1991quantized}, and Bose-Einstein condensates (BECs)~\cite{bec_review}. While optical vortices are geometric in origin, described typically by Laguerre-Gaussian solutions of the paraxial wave equation, vortices in superfluids and BECs are referred to as topological defects
with quantised circulation due to the
single-valued nature of their wavefunction.

Exciton-polariton condensates~\cite{kasprzak_bose-einstein_2006}, appearing in the strong light-matter coupling regime in semiconductor microcavities~\cite{kavokin_microcavities_2007}, lie at the interface between non-interacting optical systems and interacting quantum fluids. They can form superfluid currents at elevated temperatures~\cite{Amo2009superfluid, Amo_Nature2009, Lerario_NatPhys2017} and quantum vortices~\cite{Lagoudakis2008firtsv, Lagoudakis2009_halfquantised, Sanvitto2010_resonantvortex_superfl, Roumpos_NatPhys2011, Resonant_injection, Dominici_NatComm2018, Caputo_NatPhot2019}, and due to their strong nonequilibrium nature, populate geometric vortex states based on the balance of pump-induced gain and dissipation in confining potentials~\cite{Nardin_PRB2010, Guda_PRB2013, chiral_lens_prl, Liu_PNAS2015, Gao_vortex_chain, Ma2020_vortex_swotching, Cookson_NatComm2021, Real_PRR2021}. The salient features of exciton-polaritons (hereafter polaritons) are the extremely small effective mass due to the photon part, and large nonlinearities due to their excitonic component. The lightness of polaritons renders them excellent candidates for fundamental studies and application of polariton condensates at room temperature \cite{Plumhof2013,Daskalakis2014, Zasedatelev2019,Zasedatelev2021nature}. 
However, despite the recent progress in the field of polaritonics, to date, vortex formation in a stirred polariton condensate, as in the ``rotating bucket" experiments of liquid helium ~\cite{donnelly1991quantized} or diluted quantum gases~\cite{bec_review, bec_laser_prl1999, Abo_Science2001,Engels_PRL2003}, remains illusive predominately due to their ultrashort lifetime; limited typically to a few picoseconds. While resonant injection of orbital angular momentum into polariton fluids has been possible for some time~\cite{Sanvitto2010_resonantvortex_superfl, Resonant_injection}, the concept of dissipative superfluidity holds only under non-resonant excitation~\cite{Juggins_NatCommun2018}. Generation of polariton vorticity with deterministic direction of rotation could be possible using external electric fields inducing a pseudodrag effect~\cite{Chestnov_PRB2019}, or magnetic fields in combination with cavity TE-TM splitting~\cite{vortex_magnet_polariton}.

In this article, we realize the ``rotating bucket" experiment in a polariton condensate using an cylindrically  asymmetric in-plane optical trap induced by a composite non-resonant excitation beam that is used to inject a repulsive exciton reservoir. The excitation pattern is formed by the beating note of two frequency detuned single-mode lasers of opposite OAM that allows us to realize a dumbbell-shaped trap that rotates at ad-hoc frequencies. This composite beam is used to both create and stir the condensate through the effective torque exerted by the asymmetric shape of the photoexcited exciton reservoir. Due to the finite cavity lifetime, polaritons decay from the cavity through quasi-mode coupling to a continuum of photon states, carrying all the physical information of the condensate: density, energy, momentum, spin, and phase.  We utilise spatially-resolved interferometry to investigate the formation of topological phase defects in the condensate as a function of rotation frequency, and identify the critical conditions under which quantised vortices are formed. We observe the formation of deterministic quantised vortex states that co-rotate with the excitation pattern for a range of pump rotation frequencies. The frequency is lower-bonded to the extend to which the optically injected exciton reservoir is sufficiently populated along the circumference of the pump's rotating  profile in order for it to confine the co-rotating condensate. The higher-bound of the rotation frequency is also limited by the exciton recombination rate. When the rotation frequency exceeds the exciton recombination rate, the asymmetry of the reservoir is smeared out resulting  in a cylindrical symmetric trapping potential that does not exert any torque to the condensate.  

\section*{Results}\label{sec2}

\begin{figure}[h]
    \centering
    \includegraphics[width=1\columnwidth]{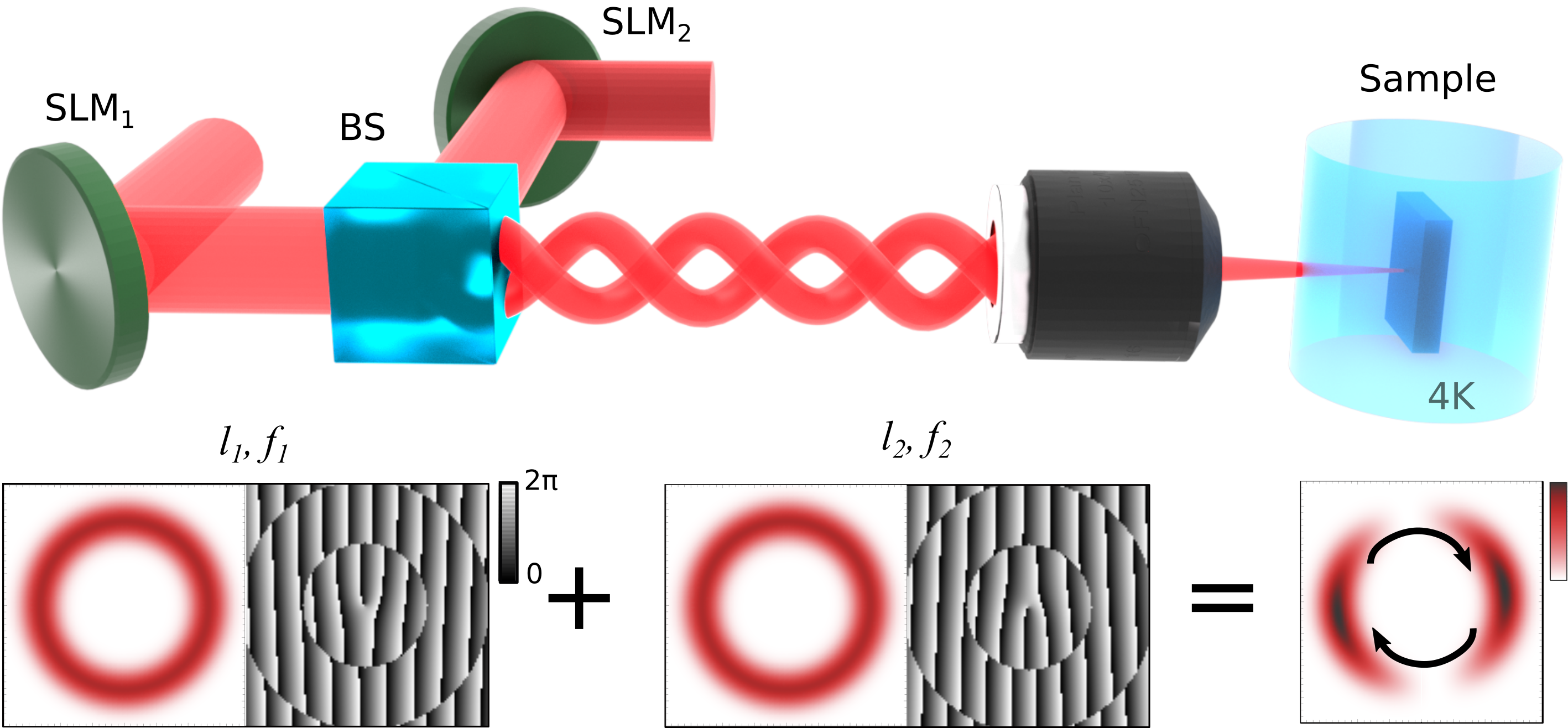}
    \caption{\textbf{ The pumping configuration in the ``rotating bucket'' experiment with polaritons.} The upper part depicts the optical excitation path of the two frequency-detuned, $f_1-f_2\neq0$, spatially modulated, single-mode lasers. The lower part illustrates the spatial intensity profile (in red color) of the two lasers that are individually shaped, with two phase-only spatial light modulators (SLMs), to form ring-like intensity profiles of opposite optical angular momentum $l_{1,2}$. The composite excitation beam acts as a rotating `dumbbell' shaped polariton trap that both injects and stirs the forming polariton condensate, similar to the optical ferris wheel for ultracold atoms. The grayscale images are the corresponding ``perfect'' vortex phase masks applied on the SLMs.}
    \label{fig1}
\end{figure}

We optically inject polariton condensate in a $2 \lambda$ inorganic microcavity consisting of two strain-relaxed, phosphor-compensated,  GaAs/AlAs$_{0.98}$P$_{0.02}$ distributed Bragg reflectors, with InGaAs quantum wells embedded at the anti-nodes of the intracavity optical field ~\cite{cilibrizzi_polariton_2014}. The polariton lifetime is $\tau_p \approx 5$ ps and the exciton-photon detuning is $-3.2$ meV. The sample is held in a closed-cycle, cold-finger cryostat at $4$ K. The excitation beam is composed of two co-circularly polarised, individually wavelength tunable, frequency-stabilized, single-mode, Ti:Sapphire lasers, that allow for precise control of their frequency detuning. Their respective wavelengths are centred around the first Bragg minimum of the reflectivity stop-band to minimise reflection losses. In Figure~\ref{fig1}, we schematically depict the excitation part of the experimental setup with the two lasers diffracting from two phase-only reflective spatial light modulators (SLMs). We apply the so-called perfect vortex mask~\cite{Chen2013_perfectvortex} on both SLMs to generate annular beam profiles with a tunable beam diameter and a rotating phase front $E_{1,2} \propto e^{i (l_{1,2} \theta - \omega_{1,2}t)}$, where $\theta$ is the microcavity in-plane angle, $\omega_{1,2}$ are angular frequencies of excitation lasers. Both applied masks are identical except for the relative sign of the OAM. We imprint opposite OAM between the two SLMs; $l_1 = \pm 1$ from SLM$_1$ and $l_2 =  \mp 1$ from SLM$_2$. Next, the two spatially modulated lasers are superimposed on a non-polarising beam-splitter and form a rotating dumbbell-shaped excitation pattern (see Fig.~\ref{fig1}). The direction and frequency of the rotation follows from the expression~\cite{rotating},
\begin{equation} \label{eqfrequency}
 f'=\frac{\Delta f}{\Delta l} = \frac{f_1-f_2}{l_1-l_2} = \frac{1}{2\pi}\frac{\omega_1-\omega_2}{l_1-l_2}.
 \end{equation}
Here, positive and negative $f'$ correspond to counterclockwise and clockwise rotation of the intensity pattern, respectively. 

The non-resonant, composite excitation pattern is projected onto the sample with an average diameter of 14 $\upmu$m. For zero frequency detuning between the two lasers, a static (non-rotating) dumbell-shaped hot exciton reservoir is injected that partially traps polaritons within the excitation profile due to the repulsive interaction between excitons and polaritons. With increasing optical excitation density we observe the formation of a polariton condensate with its centre of mass at the centre of the excitation profile, in accordance with previous studies using an annular excitation profile~\cite{askitopoulos_polariton_2013}. Due to the cylindrical asymmetry of the repulsive dumbbell-shaped exciton reservoir, here, the condensate is also spatially asymmetric as previously observed for elliptical pumping profiles~\cite{gnusov_prapl}. 

\begin{figure}[t]
    \centering
    \includegraphics[width=0.6\columnwidth]{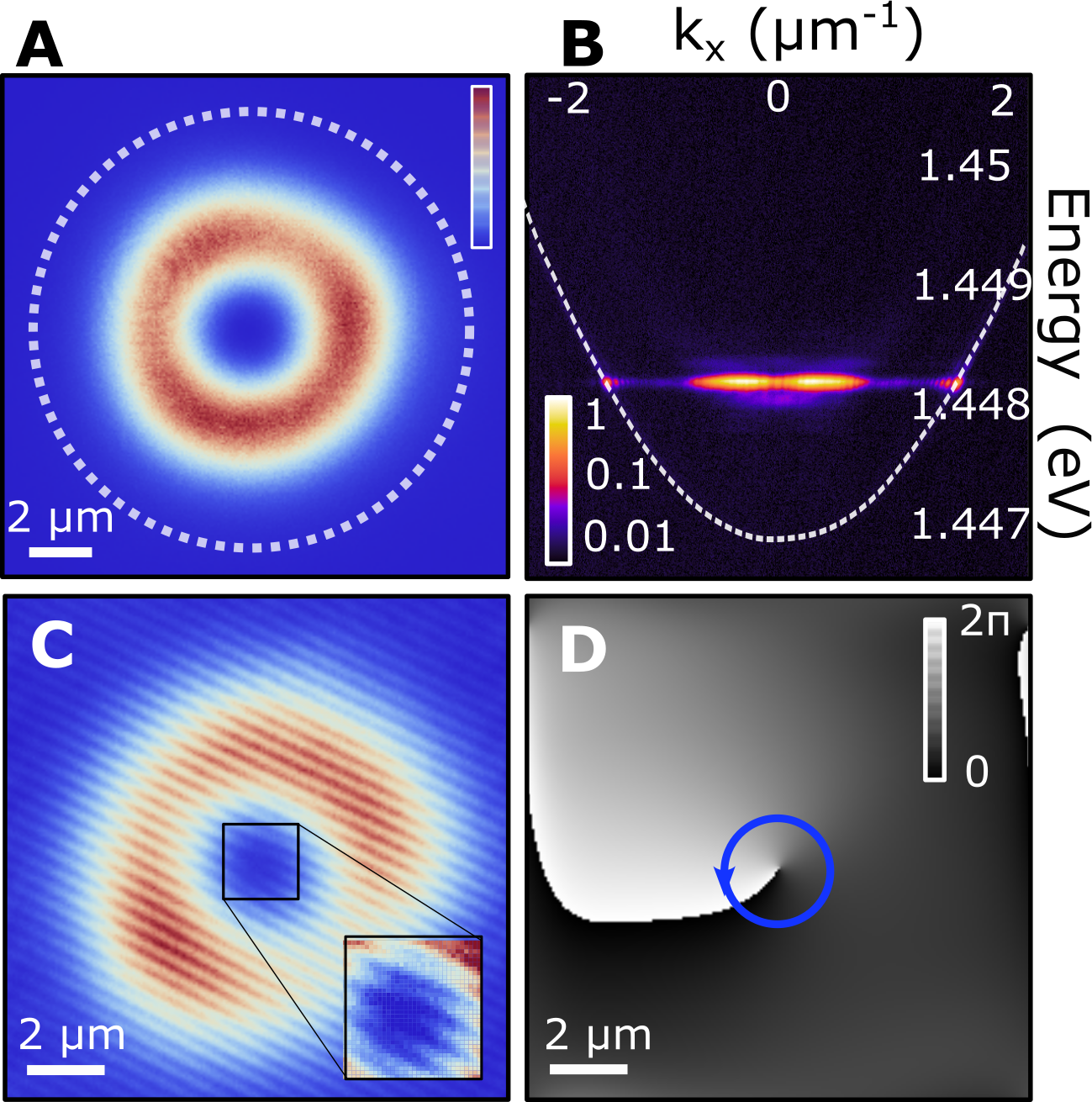}
    \caption{\textbf{ Quantised vortex formation in the ``rotating bucket'' experiment.} (\textbf{A}) Real-space normalised photoluminescence intensity of a polariton condensate under a non-resonant, counterclockwise rotating excitation beam [$\Delta f = 4.6$ GHz, $(l_1,l_2) = (1,-1)$]. The dashed line corresponds to the circumference of the effective optical trap. (\textbf{B}) Angularly-resolved normalised photoluminescence intensity of the trapped condensate (false-color in logarithmic scale). The  white dashed curve depicts the lower polariton dispersion branch. (\textbf{C}) Interference pattern of the condensate emission with a resonant, plane-wave, reference laser, revealing a fork-like dislocation in the centre of the condensate wave-function (see magnified region of interest). (\textbf{D}) Phase distribution of the condensate wave-function showing a counterclockwise winding phase singularity confirming the formation of a quantised vortex (in false-grey scale).
    }
    \label{fig2}
\end{figure}

To investigate the effect of stirring a polariton condensate by rotating the excitation pumping profile, we apply a counterclockwise rotating excitation pattern, with positive detuning between the two lasers, $\Delta f = 4.6$ GHz, opposite OAM, $(l_1,l_2) = (1,-1)$, and pumping power of $P = 1.1 P_\text{th}$, where $P_\text{th}$ is the condensation threshold pump power. Figure ~\ref{fig2}A shows time-integrated, spatially-resolved, photoluminescence measurements, wherein the dotted white circle depicts the circumference of the optical trap,  containing an annular-shaped condensate  with an intensity minimum in the centre of the pumping area. We spectrally resolve the emission in reciprocal space and observe the monochromatic characteristic of a trapped condensate (see Fig.~\ref{fig2}B)~\cite{Askitopoulos_robust}. To measure the real-space phase map of the condensate, we apply homodyne interferometry using a resonant plane reference wave~\cite{AlyatkinPrl}. Figure ~\ref{fig2}C shows the resulting interference pattern that reveals a fork-like dislocation in the centre of the annular-shaped emission pattern, implying a phase singularity. Next, we perform off-axis digital holography technique to extract the real-space phase pattern presented in Fig.~\ref{fig2}D. In the centre of condensate wave-function, at the minimum of the intensity distribution, we obtain a singularity that is indicative of a quantised vortex, with the counterclockwise phase winding  around the vortex core with OAM $l=1$; co-rotating with the excitation pumping profile. We note here that for a stationary annular trap of the same diameter we do not observe the formation of a vortex state of definite charge~\cite{askitopoulos_polariton_2013, Askitopoulos_robust}.

Our results are qualitatively reproduced through numerical modelling using a generalised 2D Gross-Pitaevskii equation describing the condensate order parameter coupled to an exciton reservoir (see Methods and Supplementary Materials section S1). 
Whilst both co- and counter-rotating geometric vortex states are solutions of the effective trapping potential generated by the excitation pumping profile, competition between gain and losses results in a quantum vortex co-rotating with the exciton reservoir. To unravel the parameter space that defines the prevalence of the co-rotating configuration, we have derived a reduced generalised Gross-Pitaevskii model, or analogously, a nonlinear Rabi flopping model in the rotating wave approximation describing the time-periodic driven dynamics of the two $l = \pm 1$ angular harmonics in the rotating potential (see Methods and Supplementary Materials section S3). The results are reminiscent of the AC Stark effect whereas in our case an asymmetric Mollow triplet appears due to losses and gain which, alongside repulsive polariton interactions, give precedence to a condensate co-rotating with the pump.

\begin{figure}[t]
    \centering
    \includegraphics[width=0.6\columnwidth]{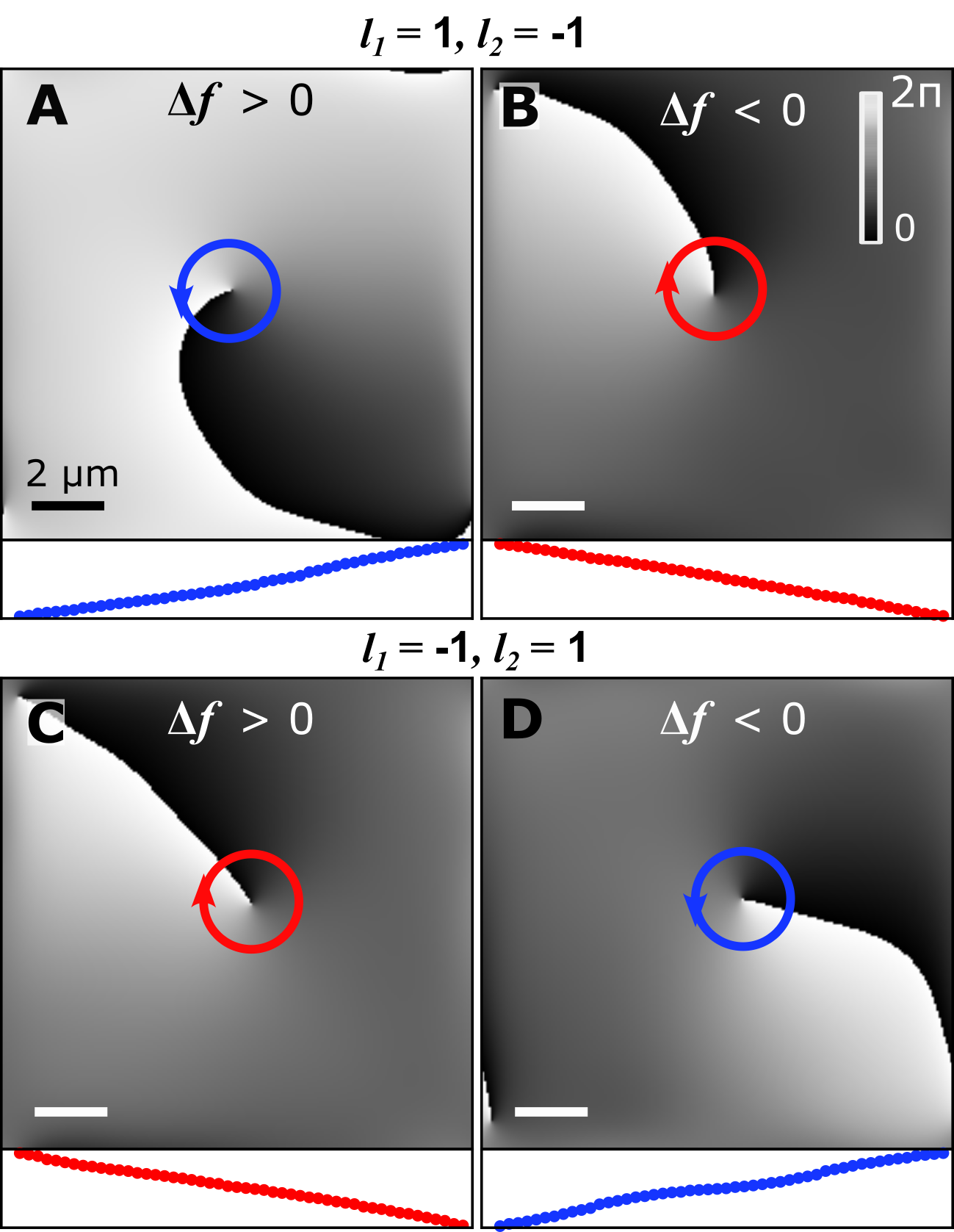}
    \caption{\textbf{Deterministic control of the quantum vortex charge.} Real-space phase distribution of the condensate wave-function demonstrating co-rotation of the winding of the vortex phase with the excitation beam. (\textbf{A} and \textbf{D}) show a counterclockwise and (\textbf{B} and \textbf{C}) a clockwise winding of the phase following the excitation beam. The winding of the phase is determined by controlling the OAM $l_{1,2}$ and lasers frequency difference $\Delta f$. In (\textbf{A} and \textbf{C}), $\Delta f = 4.6$ GHz and in (\textbf{B} and \textbf{D}) $\Delta f =-3.7$ GHz. The bottom inset of each panel (red/blue solid dots) show the nearly constant angular-phase gradient around the line-profiles of the respective phase singularity.}
    \label{fig3}
\end{figure}

Further on, we demonstrate deterministic control over the charge of the forming quantum vortex by applying the four different configurations arising from Eq.~\eqref{eqfrequency}. To control the sign of the rotation of the excitation beam, $f'$, we tune either the sign of the difference of OAM, $\Delta l$, or the sign of the frequency detuning, $\Delta f$. The real-space phase distribution in the four configurations are shown in Fig.~\ref{fig3}. The phase profiles presented in Figs.~\ref{fig3}A,B correspond to $\Delta l = 2$ and those in Figs.~\ref{fig3}c,d to $\Delta l = -2$. Figures~\ref{fig3}A,C correspond to positive detuning, $\Delta f = 4.6$ GHz, and Figs.~\ref{fig3}B,D to negative detuning, $\Delta f = - 3.7$ GHz. In all four configurations, we observe the formation of a quantum vortex, with a phase singularity at the core of the vortex and a winding of the phase co-directed with the rotation of the excitation beam with nearly constant angular-phase gradient; see circular line profiles of the phase in respective lower panels in Figure~\ref{fig3}. Beyond the interest from a fundamental studies perspective on the reproducibility of the formation of quantum vortices in rotating polariton fluids, demonstration of structured non-linear light sources with controllable topological charge offers applications in classic and quantum communications \cite{Qiu2017_science, Piccardo2022nature, Ren2016}.

\begin{figure}[t]
    \centering
    \includegraphics[width=0.8\columnwidth]{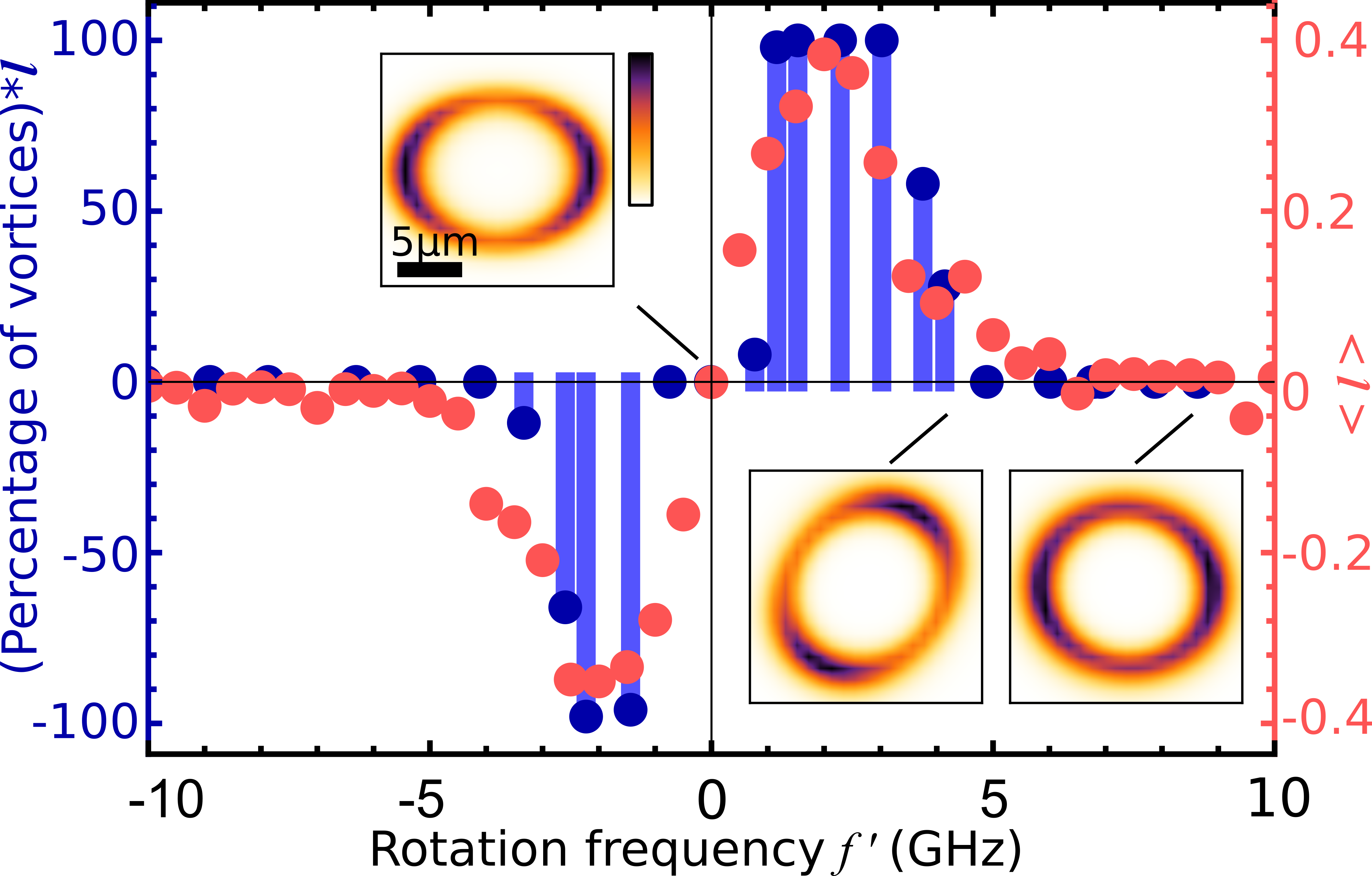}
    \caption{\textbf{Rotation-frequency dependence of a quantum vortex state formation.} Histogram of realisations of quantum vortices for $l_1=1$ and $l_2 = -1$. Blue markers show the product of the percentage of realizations wherein a quantum vortex occurs multiplied by the optical angular momentum of the resulting state at each rotation frequency $f'$. Red markers show the average angular momentum of the confined quantum vortex states obtained using a 2D generalized Gross-Pitaevskii theory. The insets depict snapshots of the exciton reservoir density distribution for 0 GHz, 5 GHz, and 10 GHz rotation frequencies.
    }
    \label{fig4}
\end{figure}

Of fundamental interest to the dynamics of the ``rotating bucket'' experiment in polariton condensates is the rotation-frequency dependence in the formation of quantum vortices. Tuning the rotation frequency of the excitation pattern with 14 $\upmu$m diameter, we observe quantum vortex formation between $\pm 1$ GHz and $\pm 4$ GHz. Due to the stochastic nature of vortex formation at the edges of this frequency range, we collect a large sample of realisations ($>10^3$) for rotation frequencies $f'$ spanning from $-10$ GHz to $10$ GHz at a constant (above condensation-threshold) excitation-density of the composite rotating beam. For each frequency, we record the interference of the real-space condensate photoluminescence with its retro-reflected, shifted image and extract the real-space phase distribution for 100 `single-shot' realisations; each, time-integrated over a $10 \ \upmu$s long excitation pulse-width (see Supplementary Materials S2). In order to distinguish quantum vortex states from other states in this large statistical sample, we develop a vortex sorting algorithm (see Supplementary Materials S2). Condensate realizations that do not qualify as vortex states in the experiment correspond typically to three scenarios: a condensate populating a Gaussian-like ground state; a condensate fractured across multiple distinct energy states; or mixture of OAM $l = \pm1$ forming a standing wave (dipole state). These cases would result in a flat, indeterminate, or step-like real-space condensate phase profiles, respectively. With blue bars, in Fig.~\ref{fig4} we plot the product of the number of instances, where a quantum vortex is observed, and the OAM of the corresponding vortex state. In agreement with the results presented in Fig.~\ref{fig3}, the flipping of the condensate vortex charge with the direction of rotation is clearly evidenced in Fig.~\ref{fig4}, where the distribution, evidently, inverts around $f'=0$. 

For small rotation frequencies, $f'\in [-1,1]$ GHz, due to the finite exciton recombination time, the exciton reservoir induced by the dumbbell-shaped excitation pattern is not sufficiently populated to build a confining repulsive potential for a quantum vortex to form, as is the case for a static excitation beam. For rotation frequencies in the range of $\pm 1$ GHz and $\pm 4$ GHz, respectively, we observe that the formation of a quantum vortex state becomes statistically significant, reaching nearly 100\% occurrence approximately at $\pm2.5$ GHz. The resonant-like occurrence of quantized vorticity, for this range of frequencies, suggests that the repulsive potential from the optically induced exciton reservoir can host an excited geometric vortex state above the trap ground state, as shown in the dispersion image of Fig.~\ref{fig2}B, and the effective stirring  starts inducing a definite direction of vortex rotation. For rotation frequencies $\lvert f'\rvert$ higher than 4 GHz, the asymmetry of the reservoir is smeared out resulting in a cylindrical symmetric trapping potential that does not exert sufficient torque to the condensate for inducing the formation of a quantum vortex state. Condensate in the case occupies the Gaussian ground state (see Supplementary Materials Section S2). The insets in Fig.~\ref{fig4} show snapshots of the modelled exciton reservoir density calculated for three different rotation frequencies, with the highest rotation frequency resulting in an almost circular profile (see Methods). In the current excitation configuration, the forging of a cylindrical symmetric trapping potential at higher frequencies prevents the formation of higher number of quantum vortices with increasing rotation frequency, as it was observed in other superfluid systems. The experimental observations are qualitatively confirmed through numerical simulations, of the 2D generalised Gross-Pitaevskii model coupled to an exciton reservoir (see Methods and Supplementary Information section S1). The numerical results for the average OAM of the confined quantum vortices as a function of rotation frequency are shown with red solid markers in Fig.~\ref{fig4}.  

\section*{Discussion}\label{sec4}

The investigation of quantum vortex formation in ultracold trapped quantum gases and liquid helium has enabled a plethora of fascinating fundamental and comparative studies of superfluids. Here, we realise and study the formation of quantum vortex states in  ``rotating bucket'' experiment of driven-dissipative quantum fluids of light based on Bose-Einstein condensates of polaritons. We emphasize that, similarly to the classical "rotating bucket" experiments ~\cite{donnelly1991quantized, bec_review}, obtained rotation frequency dependence for our system manifests both critical frequency (velocity) at which vortex appears ($f' = 1 $GHz) and the range of stirring frequencies that sustains vortex of deterministic topological charge. However, the underlying physics of polaritons shifts required stirring frequencies to GHz range (compared to sub-Hz for the superfluids and BECs) and  does not support the transition to higher topological charges. Beyond facilitating means for the fundamental study of quantum fluids of light and comparative studies with other superfluids, polariton condensates offer an alternative versatile semiconductor platform for generating nonlinear structured light. With the rapid advancements in creating extended polariton networks~\cite{AlyatkinPrl}, our method can be used to engineer vortex arrays to study the complex interplay of polarization, OAM, and linear momentum degrees of freedom in programmable large scale driven-dissipative quantum fluids. Moreover, our demonstration offers a controllable source of optical vortices that empower applications in classical~\cite{Sigurdsson_PRB2014, Ma_vvcontrol, sergey_vortex_lattice} and quantum computing~\cite{splitring,Kavokin2022}, and the potential to study non-reciprocal transport of quantum fluids~\cite{Xu_PRB2021}.

\section*{Materials and Methods}\label{sec5}

\bmhead{Generalised Gross-Pitaevskii theory}
The effects of the rotating pump profile on the condensate and exciton reservoirs are modelled using the mean-field theory approach, where the condensate order parameter $\Psi(\textbf{r},t)$ describes the condensate density as a macroscopic two-dimensional wave function following the generalised Gross-Pitaevskii equation coupled to an active and inactive exciton reservoirs $n_{A,I}(\textbf{r},t)$ \cite{wouters_excitations_2007}. The active reservoir excitons undergo bosonic scattering into the condensate whereas the inactive high-momentum reservoir excitons are not allowed to scatter into it:
\begin{gather}
i\hbar\frac{\partial \Psi}{\partial t} = \bigg[-\frac{\hbar^2\nabla^2}{2m}  + G(n_A + n_I) + \alpha\lvert\Psi\rvert^2 + \frac{i\hbar}{2}(Rn_A - \gamma)\bigg]\Psi
\label{eq:rot_GPE}\\
\frac{\partial n_A}{\partial t} = -(\Gamma_A + R\lvert\Psi\rvert^2)n_A + Wn_I
\label{eq:rot_nA}\\
\frac{\partial n_I}{\partial t} = -(\Gamma_I + W)n_I + P(\textbf{r},t).
\label{eq:rot_nI}
\end{gather}
Here, $m$ is the effective polariton mass, $G = 2g\lvert X \rvert ^2$ and $\alpha = g\lvert X \rvert ^4$ are the polariton-reservoir and polariton-polariton interactions strengths respectively, where $g$ is the exciton-exciton dipole interaction strength and $\lvert X\rvert^2$ is the excitonic Hopfield coefficient. Additionally, $R$ is the rate of stimulated scattering of polaritons into the condensate from the active reservoir, $\gamma=1/\tau_p$ is the polariton decay rate (inverse of polariton lifetime), $\Gamma_{A,I}$ are the active and inactive reservoir exciton decay rates, $W$ is the inactive to active reservoir exciton conversion rate, and $P(\textbf{r},t)$ describes the nonresonant continuous-wave dynamic pumping profile.
\begin{align} \notag
P(\mathbf{r},t) & = \mathcal{P}(r) \left\lvert e^{i (l_1 \theta -  \omega_1 t )} + e^{i (l_2 \theta -  \omega_2 t)} \right\rvert^2 \\
& = 4 \mathcal{P}(r) \cos^2{\left[ \frac{(l_1-l_2) \theta}{2} - \frac{(\omega_1 - \omega_2) t}{2} \right]},
\label{eq.pump}
\end{align}
where $\mathcal{P}(r)$ represents the annular intensity profiles shown in Fig.~\ref{fig1}. 

To understand the reservoir dynamics which play an essential role in our vortex stirring experiment, it is good to write the general solution to Eq.~\eqref{eq:rot_nI},
\begin{equation}
n_I(\mathbf{r},t) = e^{- (W+\Gamma_I) t} \left[ 4 \mathcal{P}(r) I(\theta,t) +  n_I(\mathbf{r},0)\right]
\label{eq.nIt}
\end{equation}
where the integral becomes
\begin{align}
I(\theta,t)
&= e^{(W + \Gamma_I) \tau} \frac{(W+\Gamma_I) \cos^2{(\bar{\theta} - \bar{\omega} \tau)} -  \bar{\omega} \sin{[2 (\bar{\theta} - \bar{\omega} \tau )]} + \frac{2 \bar{\omega}^2}{W + \Gamma_I}}{(W  + \Gamma_I)^2 + 4 \bar{\omega}^2} \bigg\rvert_0^t.
\label{eq.integral}
\end{align}
Here, we have simplified the notation to $\bar{\theta} =  (l_1-l_2) \theta/2$, and $\bar{\omega} = (\omega_1 - \omega_2)/2$.

In the limit of a slowly rotating trap at long times one retrieves a reservoir that adiabatically follows the shape of the pump,
\begin{equation}
n_I(\mathbf{r},t) \Big\rvert_{W+\Gamma_I \gg f'} \approx   \frac{P(\mathbf{r},t)}{W + \Gamma_I}.
\end{equation}
The same argument can be applied to the active reservoir $n_A(\mathbf{r},t)$. Assuming that nonlinear effects are weak and that the decay rate $\Gamma_A$ of the active reservoir is fast, i.e. $\Gamma_A \gg f', R \lvert\Psi\rvert^2$, then it will also adiabatically follow the dynamics of the inactive reservoir,
\begin{equation} \label{eq.nA_fast}
n_A(\mathbf{r},t) \Big\rvert_{\Gamma_A \gg f'} \approx   \frac{W n_I(\mathbf{r},t)}{\Gamma_A} = \frac{P(\mathbf{r},t)}{\Gamma_A(1 + \Gamma_I/W)}.
\end{equation}
In the opposite limit of a rapidly rotating trap ($W+\Gamma_I \ll f'$) the inactive reservoir approaches the cylindrically symmetric time-independent solution $n_{I}(\mathbf{r}) \approx 2\mathcal{P}(r)/(W+\Gamma_I)$. The same goes as well for the active reservoir.

Equations~\eqref{eq:rot_GPE}-\eqref{eq:rot_nA} are numerically integrated in time using a linear multistep method starting always from weak random initial conditions. The parameters used in these simulations are based off the sample properties~\cite{cilibrizzi_polariton_2014}, with $m = 5.3 \times 10^{-5}m_0$ where $m_0$ is the free electron mass, $\gamma = \frac{1}{5.5}$~ps$^{-1}$, $g = 1~\upmu$eV~$\upmu$m$^2$, and $\lvert X\rvert^2 = 0.35$. We take $\Gamma_A = \gamma$ due to the fast thermalisation to the exciton background and the nonradiative recombination rate of inactive reservoir excitons to be much smaller than the condensate decay rate with $\Gamma_I =0.01\gamma$. The remaining parameters are enumerated through fitting to experimental data, giving $R = 0.01$~ps$^{-1}$, and $W = 0.05$~ps$^{-1}$. The nonresonant pump drive term $P(\textbf{r},t)$ uses a similar profile as in experiment.

For each rotation frequency, $f'$, 40 
unique realisation are performed, each starting with different random initial conditions and integrated forward in time until converging to a final state. The expectation value of the OAM in the condensate plotted in Fig.~\ref{fig4} is written,
\begin{equation}
    \langle l \rangle = \frac{1}{\hbar} \frac{\langle \Psi \vert \hat{L}_z \vert \Psi \rangle }{\langle \Psi \vert \Psi \rangle}
    \label{eq:rot_avOAM}
\end{equation}
where $\hat{L}_z$ is the angular momentum operator.

\bmhead{Perturbative treatment in the linear regime}
In this section we consider a truncated Hilbert space composed of just the $l = \pm 1$ OAM modes describing the trapped polaritons in the linear regime under a time-periodic perturbation. The Schrödinger equation describing the driven system without cavity losses $\gamma$ can be written,
\begin{equation}
i \hbar \frac{\partial \psi}{\partial t} = [\hat{H}_0 + \hat{U}(t)]\psi,
\label{eq.schro}
\end{equation}
where,
\begin{align}
\hat{H}_0 &= - \frac{\hbar^2 \nabla^2}{2m} + \frac{1}{2} m \omega_\parallel^2 r^2,\\
\hat{U}(t) &= \hbar\lambda \cos^2{\left( \frac{(l_1-l_2) \theta}{2} - \frac{(\omega_1 - \omega_2) t}{2} \right)}.
\end{align}
Here, we have assumed that the polaritons are deeply enough trapped to feel an effective harmonic potential of strength $\omega_\parallel$ determined by the time-independent terms in the reservoir solutions~\eqref{eq.nIt}-\eqref{eq.nA_fast}. Here, $\lambda = \lambda_R + i \lambda_I$ is a complex number that satisfies $\lvert \lambda \rvert \ll \omega_\parallel$ (i.e., the operator $\hat{U}$ can be treated as a perturbation). This is a valid assumption for intermediate rotation frequencies when $(\omega_1 - \omega_2)^2 >2(W+\Gamma_I)^2$ where the reservoir becomes sufficiently smeared out [see Eqs.~\eqref{eq.nIt} and~\eqref{eq.integral}], so it forms approximately a sum of a time-independent confinement term $m \omega_\parallel^2 r^2/2$ and a weak time-dependent perturbating term $\hat{U}(t)$.

For $\hat{H}_0$ the solutions are 2D Hermite- or Laguerre-Gaussians modes~\cite{QMvol2}. We will truncate our Hilbert space around the degenerate pair of first-excited angular harmonics $l = \pm 1$ which are the main observation in this experiment,
\begin{equation}
\psi = \xi(r) \left( c_+  e^{i \theta} + c_- e^{-i \theta}\right) e^{-i 2 \omega_\parallel t}.
\label{eq.ansatz}
\end{equation}
Where $\xi(r) = \frac{\beta^2}{\sqrt{\pi}} r  e^{- \beta^2 r^2/2}$ is the radial solution to the unperturbed Schrödinger equation with $\beta = \sqrt{m \omega_\parallel/\hbar}$.

The angular dependence of $\hat{U}$ means that it couples harmonics that differ by $l_1 - l_2$ in angular momentum. Therefore, the case of interest $l_1 - l_2 = \pm2$ will couple together the harmonics $e^{\pm i \theta}$. It also means that the ground state is decoupled from the first excited state and therefore is safe to neglect in the expansion~\eqref{eq.ansatz}.  Plugging~\eqref{eq.ansatz} into the perturbed Schrödinger equation \eqref{eq.schro} and integrating over the spatial degrees of freedom (i.e., exploiting their orthogonality) we obtain the coupled system of equations (up to an overall energy factor),
\begin{equation}
i \frac{\partial c_\pm}{\partial t} =   \frac{\lambda}{4} c_\mp \exp{\left[\mp i(\omega_1 - \omega_2)t \frac{ l_1-l_2}{\lvert l_1-l_2 \rvert}\right]}.
\label{eq.sol}
\end{equation}
Interestingly, these are the same equations that describe Rabi flopping in a degenerate two-level system in the rotating wave approximation. This is analogous to the AC Stark effect, commonly applied in atomic optics where an oscillating electric field shifts atomic transitions. The solutions are
\begin{equation} \label{cp}
c_\pm(t) = A_\pm e^{-i(\pm\Delta \omega + \Omega_R)t/2} + B_\pm e^{-i(\pm\Delta \omega - \Omega_R)t/2}
\end{equation}
where $A_\pm,B_\pm$ are determined by initial conditions and parameters of the model, and $\Delta \omega = (\omega_1 - \omega_2)(l_1 - l_2)/\lvert l_1 - l_2 \rvert$ with $\Delta \omega<0$ corresponding to clockwise and $\Delta \omega > 0$ counterclockwise pump rotation, and $\Omega_R = \sqrt{\Delta \omega^2 + (\lambda/2)^2}$. We note that the exponents constitute the famous Mollow triplet, the hallmark of dressed quantum states. 

We can derive a useful expression which quantifies the amount of OAM in the non-Hermitian system at timescales $t\gg\mathrm{Im}{(\Omega_R^{-1})} = \nu^{-1}$. It is instructive to describe our initial condition as a vector on the unit sphere with polar and azimuthal angles $\Theta$ and $\phi$ so it reads:
$
\vert \psi(t=0) \rangle = (c_+^{(0)}, c_-^{(0)})^\mathrm{T} = (\cos{(\Theta/2)}e^{i \phi/2},\sin{(\Theta/2)} e^{-i\phi/2})^\mathrm{T}
$.
We then obtain,
\begin{align} \notag
\vert c_+(t) \vert^2 - \vert c_-(t) \vert^2  & \simeq \frac{1}{2 \vert \Omega_R\vert^2} \bigg[\left( \Delta\omega^2 -\frac{\vert\lambda\vert^2}{4} +  \vert \Omega_R\vert^2\right) \cos{(\Theta)} \cosh{(\nu t)}\\ \notag
& + \Big( \text{Im}{(\lambda \Omega_R^*)} \sin{(\phi)} \sin{(\Theta)} - 2\text{Re}{(\Omega_R)} \Delta \omega \Big) \sinh{(\nu t)}\\
& -  \text{Re}{(\lambda)} \cos{(\phi)} \Delta \omega \sin{(\Theta)} \cosh{( \nu t)}\bigg].
\end{align}

This OAM measure is similar to the Stokes parameter for circularly polarized light, except now for angular momentum. Note that when nonlinearities are included the divergence is avoided through the reservoir gain clamping down.

Averaging over all the possible initial conditions $\Theta = [0,\pi]$ and $\phi \in [0,2\pi)$---which physically represents averaging over multiple realizations of the condensate seed---we obtain, 
\begin{equation} 
\langle \vert c_+(t) \vert^2 - \vert c_-(t) \vert^2  \rangle_{\Theta,\phi}  \simeq -  \text{Re}{(\Omega_R^{-1})} \Delta \omega \sinh{( \nu t)}.
\end{equation}
This expression is interesting as it shows that without $\nu = \mathrm{Im}{(\Omega_R)}$ the OAM would vanish. Moreover, the amplitude and sign of the OAM is dictated not just by the strength and orientation of the pump $\Delta \omega$ but also the anti-Hermitian part $\nu$ (because the hyperbolic sine is an odd function). 

Our experimental observations imply that we must have $\nu<0$ in order for the condensate to be co-rotating with the pump. This means that our perturbation parameter should satisfy $\text{sgn}{(\lambda_R \lambda_I)} = -1$, which physically means that the lower energy components in our Rabi model \eqref{eq.sol} have more gain. Intuitively this makes sense because polaritons are highly interactive and relax efficiently in energy, preferentially populating low energy solutions. In section S3 in the Supplemental Material we verify this interpretation by introducing standard polariton condensate nonlinearities to Eq.~\eqref{eq.sol} and numerically solving the coupled equations of motion, averaging over many random initial conditions, for the $c_\pm(t)$ modes in the condensate.

In the special case of $\nu=0$ we get particularly simple expression in the time-average over one period $T_R = 2\pi/\Omega_R$,
\begin{equation}\label{eq.S3_simp}
\langle \vert c_+(t) \vert^2 - \vert c_-(t) \vert^2 \rangle_{T_R} \simeq \frac{\Delta\omega^2}{ \Omega_R^2} \bigg[ \cos{(\Theta)} -  \frac{\lambda}{2\Delta \omega} \cos{(\phi)} \sin{(\Theta)} \bigg]
\end{equation}

Notice that if one averages the OAM also over all initial conditions on the unit sphere then the above expression is zero. This underlines the crucial difference in the OAM dynamics between the Hermitian ($\nu=0$) and non-Hermitian ($\nu\neq0$) case.

Post-submission note: the following manuscript was submitted to the arXiv describing an alternative technique for the non-resonant rotation of a polariton condensate~\cite{arxiv.2209.01904}.

\section*{Acknowledgments}

I.G. acknowledges the helpful discussions with Daniil Pershin.

The authors acknowledge the support of the European Union’s Horizon 2020 program, through a FET Open research and innovation action under the grant agreement No. 899141 (PoLLoC) and No.964770 (TopoLight) and UK’s Engineering and Physical Sciences Research Council (Grant No. EP/M025330/1 on Hybrid Polaritonics).

S.A.  and P.G. acknowledge the financial support of the Russian Foundation for Basic Research Grant No. 20-52-12026.

H.S. acknowledges the Icelandic Research Fund (Rannis), grant No. 217631-051.

S.H. acknowledges the use of the IRIDIS High Performance Computing Facility and the associated support services at the University of Southampton.

\section*{Contributions}

I.G., S.A. and K.S. performed the experiments and analysed the data.  J.T. developed the software for the data acquisition. S.H. and H.S. developed the
theory and carried out numerical simulations. I.G. and P.L. designed the experiment. P.L. supervised the project. All the authors participated in the writing of the manuscript.

 \section*{Competing interests.} 

Authors declare that they have no competing interests.    

\section*{Supplementary materials}
Supplementary Text\\
Figs. S1 to S3

 \section*{Data and Materials Availability.} 
 
All data supporting this study are openly available from the University of Southampton repository~\cite{repsitoriy}.


\setcounter{equation}{0}
\setcounter{figure}{0}
\setcounter{section}{0}
\renewcommand{\theequation}{S\arabic{equation}}
\renewcommand{\thefigure}{S\arabic{figure}}
\renewcommand{\thesection}{S\arabic{section}}

\newpage
\vspace{1cm}
\begin{center}
\Large \textbf{Supplementary Information}
\end{center}

\section{Condensate intensity and phase distribution simulated with 2D generalised Gross-Pitaevskii equation.}

\begin{figure}[h]
    \centering
    \includegraphics[width=0.7\columnwidth]{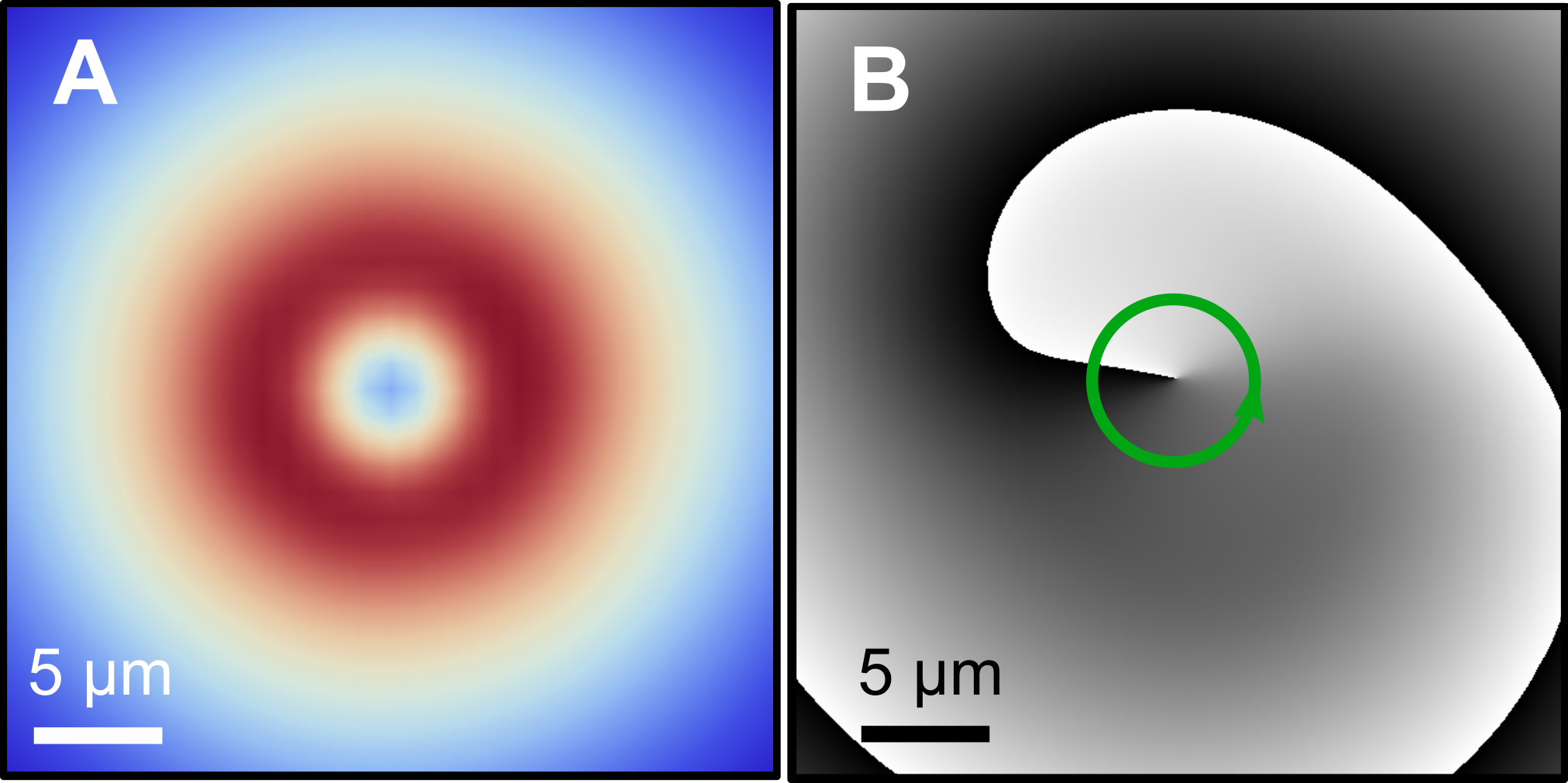}
    \caption{\textbf{Condensate intensity and phase distribution simulated with 2D generalised Gross-Pitaevskii equation.} Simulated  time integrated real-space intensity \textbf{A} and instantaneous phase \textbf{B} of the condensate at $f'=$ 2 GHz demonstrating the formation of the quantised vortex after 800~ps of numerical integration. 
    }
    \label{fig.sim}
\end{figure}

Our simulations using the generalized Gross-Pitaevskii theory (presented in main manuscript) reveal that the resulting condensate, driven by the rotating excitation profile, forms a similar time-averaged annular density structure (presented in Fig.~\ref{fig.sim}A) to the experimental findings shown in the main text (Figure~2), while possessing the vortex phase singularity (see Fig.~\ref{fig.sim}B) and phase winding co-rotating with the excitation pumping profile.

\section{Polariton interferometry and Vortex sorting procedure}

We retrieve the phase of the polariton condensate utilizing two interferometric techniques: the homodyne interferometry technique~[41], and the more common interference of the condensate emission with a retro-reflected copy of itself in a Michelson interferometer~[12].

The homodyne technique is described in detail in~[41]. We use an external diode laser with the same energy as the condensate as a reference wave and interfere it with the condensate emission. The reference laser beam is greatly expanded to spatially overlap with the entire collected photoluminescence pattern in order to have a flat phase front in the vicinity of the condensate. To lock the phases between the reference laser and the condensate, we use a small portion of the reference laser to locally seed the condensate phase. This allows for the observation of high-contrast interference fringes (see Fig.~2C in the main text) and for the phase retrieval of the condensate.

\begin{figure}[h]
    \centering
    \includegraphics[width=0.7\columnwidth]{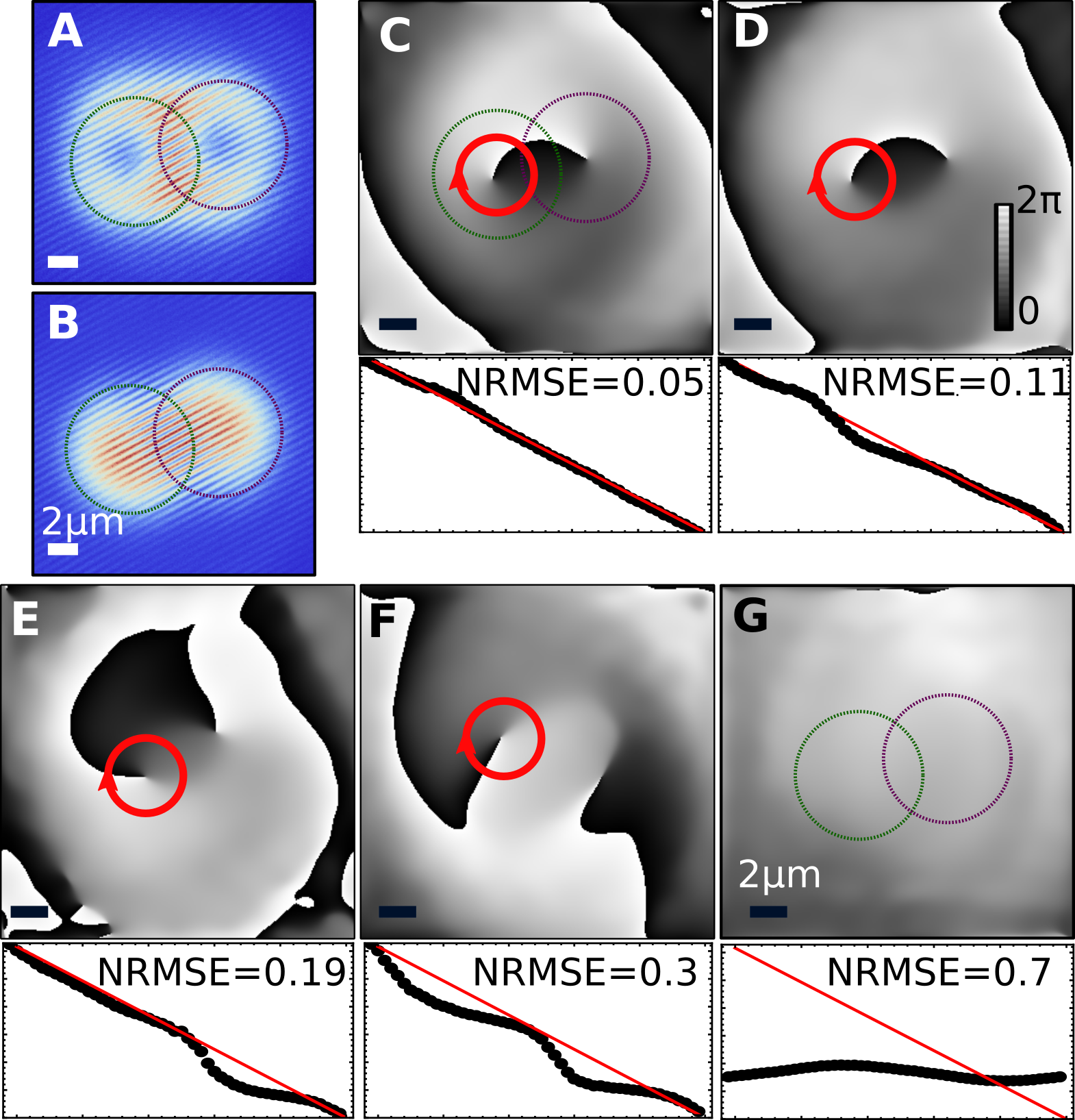}
    \caption{\textbf{Examples of polariton condensate phase distribution under different stirring frequencies.} Condensate emission (green ring) interfered with the retro-reflected and displaced copy of itself (purple ring) at $f^\prime = - 2.5$ GHz (\textbf{A}) and $f^\prime = - 8.2$ GHz (\textbf{B}). Note, that green and purple circles in (\textbf{A}-\textbf{C}) schematically depict the maximum of the condensate intensity distribution in Figure a for direct and retro-reflected copy respectively. The less bright and out-flowing part of the condensate is also present and results into the interference fringes outside the rings.  (\textbf{C}-\textbf{G}) Examples of the condensate phase distribution for different realizations with a 10 $\mu$s excitation pulse width. The bottom panels of (\textbf{C}-\textbf{G}) represent the annular line profile with a 2.6 $\mu$m radius around each phase singularity unwound starting from the black-white colour dislocation and compared to the phase profile of a perfect antivortex (red line). In panels (\textbf{C} and \textbf{D}) $f'= -1.5$ GHz, and in (\textbf{E} and \textbf{F}) $f' = -3.8$ GHz, and in (\textbf{G}) $f' = -8.2$ GHz.
    }
    \label{fig.profile}
\end{figure}

For the interference measurements of the condensate with itself used to obtain the data presented in Fig.~4 in the main text, we do not use an external light source, but instead we spatially displace one arm of the interferometer with respect to the other (retroreflected) to observe the phase dislocation. As a result, when a vortex is present in the condensate, we see instead a vortex-antivortex pair in the interference pattern due to the angular momentum flipping caused by the retroreflection (see Fig.~\ref{fig.profile}).

Given the interference pattern, one can extract the phase of the condensate wave function by performing the 2D Fourier transformation. In the frequency domain, we filter the harmonic corresponding to the fringes period and do the inverse Fourier transform. The resultant complex-valued array carries the information about the phase of the initial distribution, which we can retrieve by taking the complex argument of the values in this array.

 Example of  interference images, used to obtain the data in Fig.~4 in the main manuscript, are presented in Figs.~\ref{fig.profile}A,B. Figure~\ref{fig.profile}A corresponds to the $f' = -2.5$~GHz rotation frequency and depicts the forklike dislocations corresponding to the phase singularity of the condensate vortex. Note the two counter-directed forks in the interference pattern Fig.~\ref{fig.profile}A, one of which directly corresponds to the vortex present in the condensate 
and the other to the spatially shifted and retroreflected condensate signal with flipped optical angular momentum (OAM). Figure~\ref{fig.profile}B corresponds to the condensate occupying the Gaussian ground state of the trap at high rotation frequency $f' = -8.2$ GHz. The corresponding phase profile reveals a flat phase shown in Fig.~\ref{fig.profile}G with no OAM. 

Figures~\ref{fig.profile}C-F depict the phase distribution for different condensate realizations at different rotation frequencies (Fig.~\ref{fig.profile}C,D and Fig.~\ref{fig.profile}E,F correspond to  $f' = -1.5$ GHz and $f' =- 3.8$ GHz, respectively). The circular line profiles around the phase singularities are presented in the lower panels. The difference between the ideal antivortex profile (see the red line) and the experimental phase profile (black dots) appears due to various reasons. First, overlapping the condensate to itself leads to the interference of two anisotropic (non-flat) phase fronts resulting in phase scrambling. Second, the condensate can simultaneously occupy neighbouring energy states (ground and excited states of the trap) which can lead to a mixture of phases corresponding to specific energy states during phase retrieval procedure. The difference between ideal and measured winding in the angular phase profile defines a true/false threshold on whether a vortex exists in the condensate or not.

\textit{Vortex sorting algorithm.} For the histogram presented in Fig.~4 in the main manuscript, we take the threshold value of the normalized root mean square error (NRMSE) to the ideal vortex as $\leq$ 0.2 when defining whether we observe a vortex or not. See, for example, Fig.~\ref{fig.profile}E for typical phase distribution with extracted value of NRMSE = 0.19 close to the threshold error value. Extracted value NRMSE $=1$ means maximum error where all phase datapoints deviate by $\pi$ radians from the expected profile.

\section{Solution of reduced generalized Gross-Pitaevskii equation }
\begin{figure}[h]
    \centering
    \includegraphics[width = 0.99\linewidth]{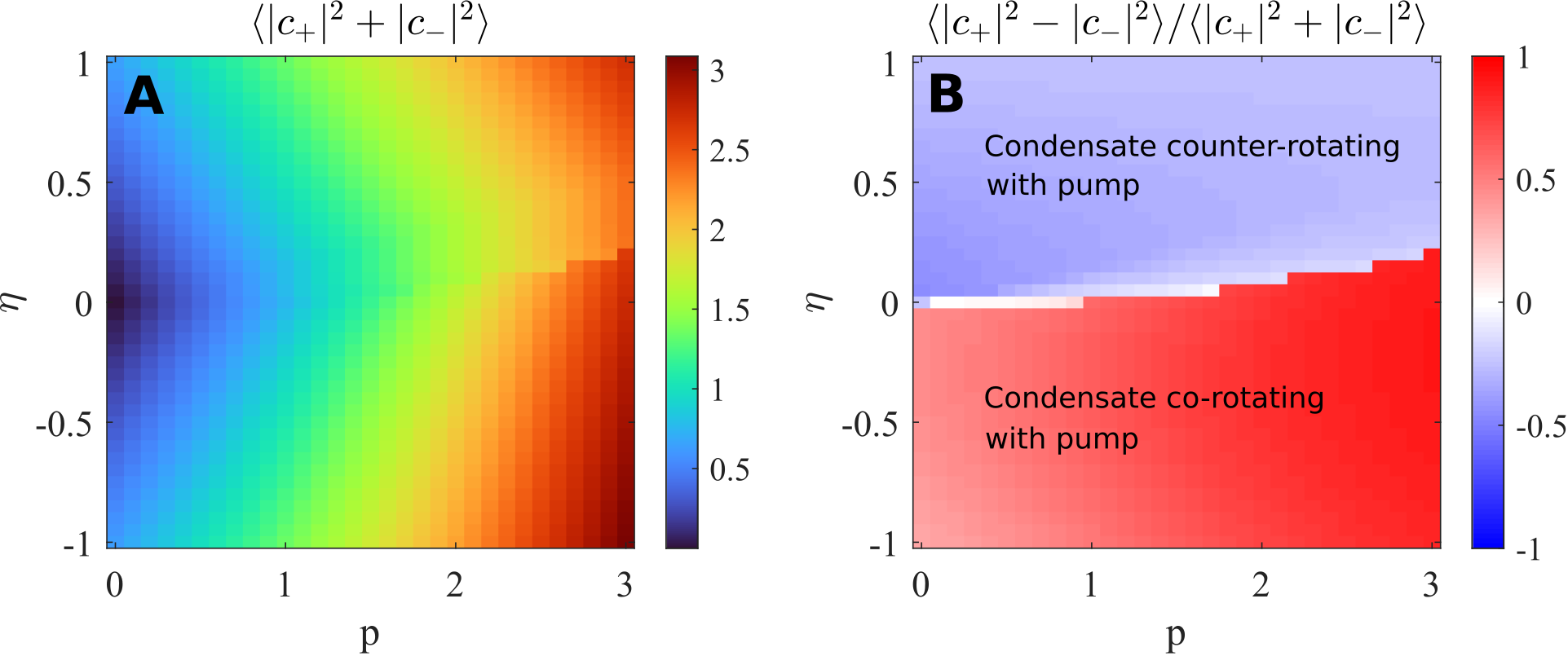}
    \caption{\textbf{Results of the numerical simulation of the Reduced Gross-Pitevskii equation.}\textbf{A} Time-averaged intensity and \textbf{B} vorticity of the condensate by numerically solving Eq.~\eqref{eq.dGPE} as a function of power $p$ and dissipative coupling strength $\eta$. Other parameters are $\tilde{\alpha} = 1$ and $\Delta \omega = 1$ (pump rotating counterclockwise).}
    \label{fig5}
\end{figure}

As an extension to our theoretical analysis presented in the main manuscript, we write a reduced generalised Gross-Pitaevskii model by introducing the standard polariton condensate nonlinearities to Eq.~(15), 
\begin{equation}
i  \frac{dc_\pm}{d t} =  \left[i p + (\tilde{\alpha} - i) ( \lvert c_\pm \rvert^2 + 2 \lvert c_\mp \rvert^2 )\right] c_\pm + (1 + i \eta)c_\mp e^{\mp i\Delta \omega t}.
\label{eq.dGPE}
\end{equation}
As discussed around Eq.~(14) in the main text, $c_\pm = \sqrt{N_\pm}e^{i \phi_\pm}$ are complex order parametersr that describe the amount of polaritons in each OAM $l = \pm1 $ modes and their phase. Here, $p$ is a denotes the difference between the pump gain and linear polariton losses, $\tilde{\alpha}$ describes the renormalized nonlinear energy shift after integrating out the spatial degrees of freedom. We have also scaled time in units of conservative coupling strength $t \to t/\lambda_R$ and defined $\eta = \lambda_I/\lambda_R$. Note that from our analysis in the main manuscript, we expect a condensate solution co-rotating with the pump when $\text{sgn}{(\nu)}=\text{sgn}{(\eta)}=-1$. The results on the time-average $\langle \dots \rangle$ behaviour of the coupled system (in the long time limit) are presented in Fig.~\ref{fig5}, also averaged over many random initial conditions, which show that depending on the sign of the dissipative coupling strength $\eta$ the vorticity can flip with respect to the pump. Here we have set $\tilde{\alpha} = 1$ and $\Delta \omega = 1$ which corresponds to repulsive polariton interactions and counterclockwise rotating pump pattern, respectively. If we flip the sign of $\Delta \omega$ then the colorscale in Fig.~\ref{fig5}B also flips. Note the slight bending of the boundary separating blue and red regions which can be attributed to the nonlinear energy term in Eq.~\eqref{eq.dGPE} becoming comparable to the dissipative coupling rate $\eta$, favoring a co-rotating polariton fluid. If the sign of $\tilde{\alpha}$ is reversed to become negative (i.e., focusing Kerr effect) the boundary will bend the other way. These results evidence that the repulsive polariton interactions favor a co-rotating polariton fluid like we observe in our experiment. 

\end{document}